# Empowering IoT Security: On-Device Intrusion Detection in Resource Constrained Devices

Vasilis Ieropoulos, Eirini Anthi, Theodoros Spyridopoulos, Pete Burnap, Aftab Khan and Pietro Carnelli
[1]Cardiff University, School of Computer Science & Informatics, Cardiff, UK
[2] Toshiba Bristol Research and Innovation Laboratory, 32 Queens Square, Bristol, BS1 4ND, UK

***Abstract*—IoT devices particularly microcontrollers are challenged by their inherent limitations in processing capabilities, memory capacity, and energy conservation. Securing communication within IoT networks is further complicated by the heterogeneity of devices and the myriad of potential security threats. Our study introduces a lightweight model that utilises machine learning algorithms to achieve a notable detection accuracy of 99% using a decision tree method and 96% using a neural network in identifying cyber threats, including Denial of Service and Man-in-the-Middle attacks which make up the majority of the attacks these devices face. While the decision tree method offers higher accuracy, it requires more computational resources, whereas the neural network approach, despite a slightly lower accuracy, is more memory-efficient. Both methods enhance the real-time monitoring and defence of IoT networks, safeguarding the transmission of data. Additionally, our approach is tailored to conserve memory and optimise computational demands, rendering it suitable for deployment on microcontrollers with limited resources.**

***Index Terms*—Internet of Things (IoT), Machine Learning, Resource Constrained Devices**

## I. Introduction

THE Internet of Things (IoT) is characterised as a network comprising physical entities, or "things," integrated with sensors, software, and other technological innovations. The primary objective of this integration is to facilitate connectivity and data exchange with other devices and systems via the Internet. This network encompasses a broad spectrum of devices, from commonplace domestic items to advanced industrial instruments. The rise of IoT has led to a surge in microcontroller-based devices that offer low cost and power consumption with ease of deployment. However, their limited processing power and memory make them susceptible to cyberattacks [1]. Implementing Intrusion Detection Systems (IDS) on these devices could provide autonomous protection against security threats [2].

In this paper, we assume a typical home environment where the user lacks the technical knowledge to deploy advanced network security solutions like IDS and firewalls. This scenario is very common, as many users do not even know how to change their router settings, often leaving them at the default. This widespread issue has prompted regulatory actions, such as the European Union's mandate against the use of generic simple passwords [3].

It is important to note that our study does not propose a generalised solution but rather a feasibility study to assess the impact of implementing an IDS on a resource-constrained device and its overall functionality. The most commonly employed practice for network security is to implement a network-wide centralised IDS, which allegedly provides security for the whole network. However, these conventional methods may not always suffice to protect devices characterised by resource constraints and real-time operational demands [1]. This reliance on a third party creates a single point of failure; if the third-party IDS is compromised, it could lead to network downtime, leaving the rest of the network vulnerable to further attacks.

Traditional IDS systems also face challenges in monitoring data in segmented networks, a common practice for IoT. This segmentation creates a layer of abstraction, making it difficult for IDSs to accurately identify anomalous behaviour, potentially leading to unauthorised access and data manipulation, thereby compromising the integrity and confidentiality of information within these networks. Consequently, hosting an IDS directly on microcontroller-based devices presents an appealing solution, allowing for independent threat detection and response without sole dependence on external security components [4].

The emergence of advanced microcontrollers, exemplified by platforms like the ESP32 with its dual processing cores, creates the potential for more sophisticated security features, including the implementation of an onboard IDS [5]. This dual-core architecture provides the computational capacity needed to execute intricate intrusion detection algorithms and respond to potential threats promptly. However, integrating an onboard IDS on microcontroller-based IoT devices presents challenges. Even with advances like the ESP32, limitations in memory and processing power affect IDS efficiency [1] [4].

While research such as that by Rizvi et al. [6] or Sforzin et al. [7] has explored the security of resource-constrained devices, it has mostly focused on single-board computers capable of running full-fledged operating systems. Our research diverges by implementing an IDS on truly resource-constrained IoT devices operating on bare metal. To the best of our knowledge, this is the first paper to investigate the deployment of an Intrusion Detection System using machine learning on such devices, aiming for faster attack response, minimising downtime, and simplifying rogue device identifi-cation. If implemented, these techniques could reduce reliance on external devices for network monitoring in segmented networks, thus reducing maintenance and deployment costs.

[0]The Authors Vasilis Ieropoulos, Eirini Anthi, Theodoros Spyridopoulos, and Pete Burnap are with Cardiff University, School of Computer Science & Informatics, Cardiff, UK. Aftab Khan and Pietro Carnelli are with Toshiba Bristol Research and Innovation Laboratory, 32 Queens Square, Bristol, BS1 4ND, UK. Corresponding Author: Vasilis Ieropoulos (email: ieropoulosv@cardiff.ac.uk)

This paper explores the potential and challenges of embedding an Intrusion Detection System (IDS) directly into microcontroller-based IoT devices. It highlights the advantages of such an approach, including device autonomy and enhanced security measures, while also acknowledging the practical constraints imposed by limited resources. This comprehensive analysis aims to provide valuable information on a critical aspect of IoT security, thus contributing to the dynamic discourse on IoT cybersecurity.

The paper is structured as follows: In Section II, we examine the current issues plaguing these devices and analyse the most prevalent attacks they are prone to. In Section IV, we elaborate on the methodology of our experiments, their results, and the trade-offs associated with each approach, providing a detailed account of our experimental journey and insights. Section V discusses the limitations of our approach, offering a balanced perspective on the scope and applicability of our methods. Finally, in Section VI, we explore potential future directions, indicating the scope for further research and improvement. This section bridges future explorations in this field, highlighting the dynamic and evolving nature of IoT security.

## II. Background

Globally, the IoT exerts a profound influence on both residential and commercial sectors. In 2022, the worldwide market for smart homes was estimated at $80.21 billion and is anticipated to expand from $93.98 billion in 2023 to $338.2 billion by 2030 [8]. By the same year, the IoT is expected to generate economic value ranging from $5.5 trillion to $12.6 trillion worldwide, inclusive of the benefits accrued by consumers and businesses utilising IoT products and services [9]. The economic potential of the IoT is predominantly concentrated in specific environments where it is implemented. For example, it is projected that the industrial sector will capture the largest share of the economic value of IoT, approximately 26 %, by 2030. Following closely, the healthcare sector is estimated to represent approximately 10 to 14 % of the economic value of the IoT by the same year. Nevertheless, the proliferation of IoT has concurrently escalated cybersecurity risks [10] [11] [12]. IoT devices are especially susceptible to network incursions, including data theft, phishing, spoofing, and distributed denial of service (DDoS) attacks. Such vulnerabilities can precipitate further cybersecurity concerns, such as ransomware and severe data breaches, which can impose substantial financial and operational burdens on affected enterprises. Recent years have witnessed a series of notable cybersecurity incidents within the IoT domain. For example, in 2021, IoT attacks increased by 50 % in a mere six-month period, with researchers finding more than 1.5 billion IoT attacks, highlighting the growing threat landscape [12]. The Apache Log4j Vulnerability, known as Log4Shell [13], impacted an extensive array of devices and was deemed one of the gravest security flaws in recent times, affecting hundreds of millions of devices. In November 2023, Infosys, an Indian IT services corporation [14], experienced a “security event” that rendered numerous applications in its U.S. division inaccessible [15], demonstrating the far-reaching consequences of such breaches. Additionally, in the same month, a substantial data breach exposed health-related data, including Covid test results, of approximately 815 million Indian citizens [16], underscoring the criticality of data protection in the healthcare sector [17]. These occurrences underscore the critical need for the implementation of stringent security protocols to protect IoT devices and networks against cyber threats.

### A. *Microcontrollers in IoT*

Microcontrollers play an important role in the Internet of Things (IoT) landscape, providing essential computational capabilities, memory, and input/output peripherals [18].

In the realm of smart home systems, microcontrollers contribute significantly to the automation of various functions, thereby enhancing comfort, convenience, and energy efficiency. For instance, Arduino boards, such as Arduino Uno, Arduino Mega, and Arduino Nano, are frequently employed because of their rich input/output interfaces and ease of programming. These boards are suitable for a wide range of smart home applications. Additionally, ESP8266 and ESP32, low-cost, low-power Wi-Fi modules that integrate microcontrollers with Wi-Fi communication functions, are extensively used in smart home devices such as smart sockets, smart bulbs, and smart switches. STMicroelectronics offers multiple series of microcontroller products, such as the STM32 series and the STM8 series. These microcontrollers, characterised by high performance, low power consumption, and rich peripheral interfaces, are suitable for controlling and connecting smart home systems [18]. In Industrial Automation, microcontrollers are utilised for the monitoring, control, and automation of production lines. The selection of microcontrollers over microprocessors in IoT applications is primarily due to their small size, low power consumption, high integration characteristics, and their ability to provide sufficient computing power. This choice also helps in maintaining low costs, complexity, and energy usage. They are available in different bit options, with 8-bit microcontrollers being used in cost-constrained but more elaborate applications than the 4-bit microcontroller [18].

### B. *IoT layers and Components*

Drawing on the OSI model’s seven-layer framework [19], IoT similarly adopts a layered architecture to facilitate its development and deployment. However, unlike the OSI model, which received widespread consensus from numerous organisations, the layered framework for IoT lacks a universally accepted standard. In our case, we shall explore the IoT layers as proposed by Cisco, IBM and Intel which consists of 7 layers [20]. As we can observe in Figure 1 the layer starts with the physical Devices and controllers. These are components like sensors, edge nodes and devices that sit on the edge that gather telemetry ready for processing. Moving up, layer 2 is made up of processing and communication units such as ESP32 [21] and ESP8266 [22]. Layer 3 consists of the Edge computing element, which is in charge of analysing the data and transforming them in a readable way. Layer 4 is responsible for storing and managing the data, usually consisting of network or cloud storage devices. Layer 5 is responsible for managing access to the data and devices through some form of access control system. Layer 6 consists of dashboards, gateways, and a control system which are in charge of bringing everything together to the end user. Finally, the top layer of the stack is the collaboration and process layer, which is outputted to the end user.

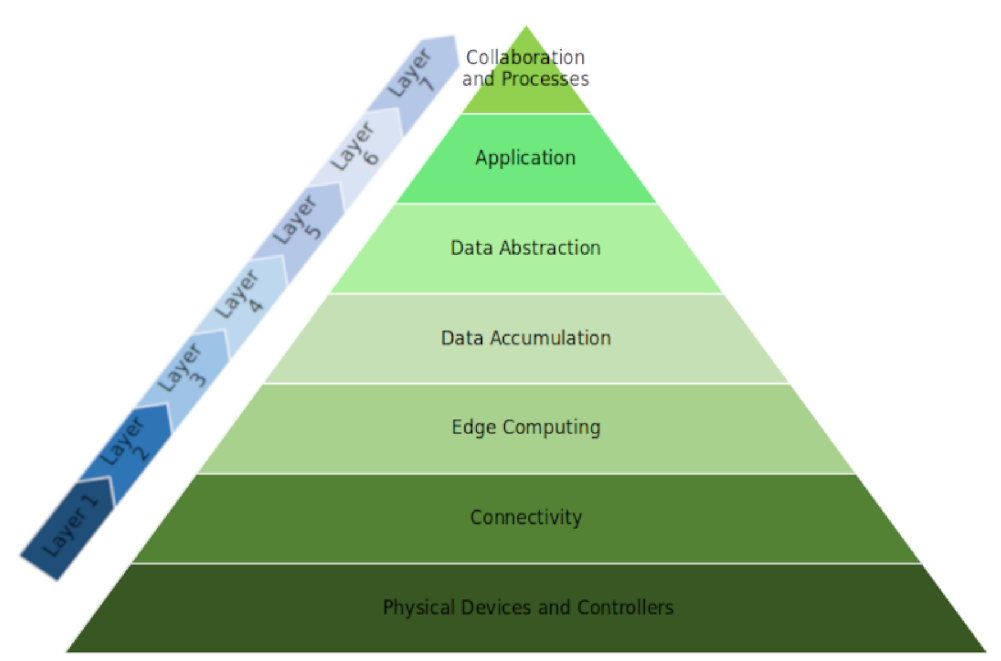

**Fig. 1:** IoT Layer Diagram

### *C. Attacks in different Layers*

The classification of the different elements of the IoT into layers, as presented in figure 1 , allows us to analyse the possible vulnerabilities and attacks that could occur in each layer. By looking at different sources we can observe that sometimes the classification of the attack varies hence we have combined the conclusions from different sources into one table [23] [24] [25]

In table I we have outlined the attacks in each layer.

| Layer | Elements | Possible attacks |
|---|---|---|
| 1 | Sensors, Motors , Edge Nodes | Malware, RE, BF |
| 2 | Communication chips, Processors | MITM, BF, Replay , Sniffing |
| 3 | AI/ML Data analysis | Data Manipulation |
| 4 | Local/NAS/Cloud storage | Ransomware, Data Manipulation |
| 5 | NAC/ACS | Phishing , SQL injection , DNS poisoning , BF |
| 6 | Web, Internet, Data centers | DDOS, Malware |
| 7 | Cloud computing , Business applications | Side channel , Malware , MITM DOS |

**TABLE I:** Layers vs Attacks

To further analyse the severity of the attacks, we classify the attacks into different layers so that we can observe which are the most frequent attacks on each layer. Concerning Figure 2 which illustrates the different attacks per layer, Layers 1, 2, and 7 seem to be the most vulnerable to attacks, as most datasets seem to focus on these 3 layers, as we can see in Table II and analyse further in subsectionII-D. However, it should be noted that not all devices cover all layers. For example, a weather station consisting of sensors and transmitters will encompass only layers 1 and 2. Furthermore, devices which operate on Layer 7, consist of Cloud services and high-end computers which theoretically already have multiple security measures implemented. Since the paper aims to focus on low-end devices such as smart metres, single-board computers, etc., it makes sense to focus only on layer 1 and 2 defence techniques.

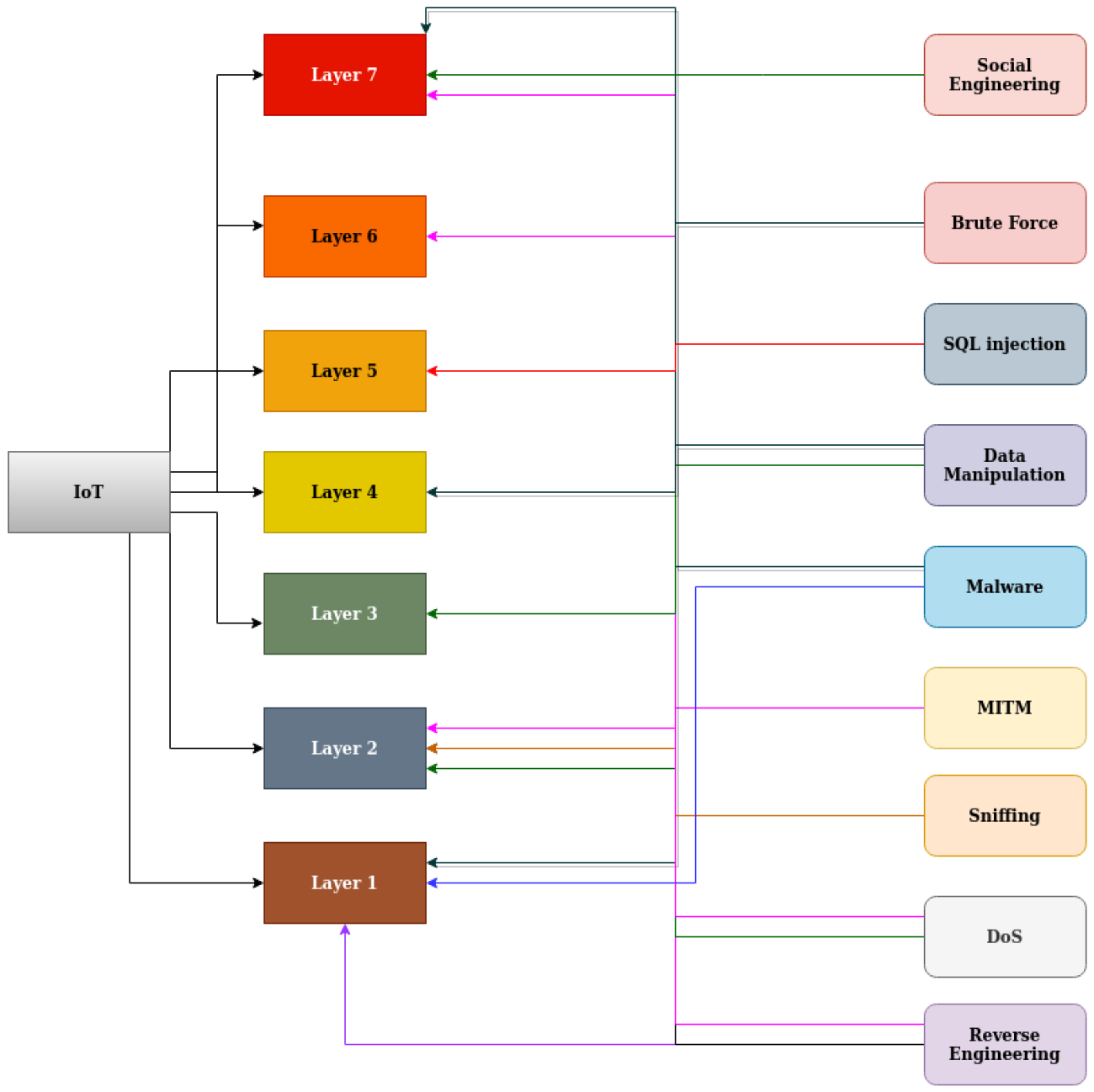

**Fig. 2:** Types of Attacks Relevant to each layer

### *D. Datasets and attacks*

Considering Figure 2, it helps us understand which attacks are most relevant and highlights. Most attacks target the devices we're examining, especially focusing on vulnerabilities at Layer 1 and 2. A survey by Ahmed Alshaibi et al. [26] provides valuable insights, as shown in Table II which shows that most established datasets mostly focus on DOS, Sniffing and man-in-the-middle attacks. After an analysis of the most popular datasets used in IoT, it is clear that DoS attacks are the most prevalent. This observation aligns with our findings from the ISO layer analysis. DoS attacks, which aim to make a machine or network resource unavailable by overwhelming it with traffic, pose a significant threat to IoT devices. Furthermore, these datasets have not included Man-in-Middle attacks or sniffing attacks, which are also possible attacks that arise in Layers 1 and 2. Hence, it is evident that a new dataset is to be created which features these types of attacks.

*1) Examples of Attacks:* There are a few notable real-world examples which have used Denial of Service and man-in-the-middle attacks to exploit vulnerable IoT. During a 2021 threat report by IBM [38] it was deduced that for the year 2021, an increase of 500% of IoT-related attacks was detected. Most of the compromised devices were poorly secured and have been infected with the Mozi and Mirai variants of botnets [38]. The study has also shown that 89% of IoT attacks detected in 2020 were infected with the Mozi botnet. The country most affected by this attack was Japan, with 80.53% of botnet attacks originating from Japan [38]. One similar but not malicious attack was executed in 2018 as a form of Hacktivism, where more than 50,000 printers were exploited to print out spam messages [39]. As analysed by Asreen Rostami et al [40], the ”PewDiePie hack” emerged from a YouTube subscriber battle, involving hacks targeting printers and video players like Chromecast. Reddit discussions among affected users highlighted various motivations behind the hack. Notably, hackers appeared to lack specific targets or vendettas, instead exploiting a router feature, Universal Plug and Play, to raise awareness about its vulnerability. Some hackers seemed motivated by a desire to showcase their skills or draw attention to specific YouTube channels. Similar discussions arose regarding printer hacks by TheHackerGiraffe, aiming to expose vulnerabilities in network printers to a wider audience. While hackers were perceived both negatively and positively, some saw their actions as highlighting significant security issues that warranted attention and resolution. In this case, the malicious actor sent ”Subscribe to PewDiePie” to thousands of printers by exploiting driver vulnerabilities, causing them to print the text. Although the attack was not malicious, it did expose a vulnerability hidden inside the printer firmware, which led to the development of patches and solutions for the issue. [39]

**TABLE II:** Comparison of Datasets

| Reference | ML Technique | IoT Attacks | IoT Dataset |
|---|---|---|---|
| [27] | OS-ELM | Dataset Multiple | NSL-KDD |
| [28] | NN | DOS, U2R and R2L. | DOS, U2R and NSL-KDD |
| [29] | DT and NB. | NNProbing, U2R and R2L. | R2L. NSL-KDDNSL-KDD |
| [30] | TAB | DoS Flooding | KDD99 |
| [29] | DT | DTandNB.DOS, Reconnaissance U2R,R2L., Backdoor | NB.Probing, U2R and R2L. NSL-KDDKDD99 |
| [31] | Ensemble Learning | Malware | AndroZoo, Drebin |
| [32] | DT. | DOS | DOS, Reconnais-RPL-NIDDS17 |
| [28] | DT | DOS, Reconnaissance U2R, R2L. | UNSW-NB15 |
| [33] | NN. | DTProbing, U2R and R2L | sance U2R, R2L., NSL-KDD |
| [28] | DT | DOS, Reconnaissance U2R, R2L. | NSL-KDD |
| [34] | NN. | DOS , Reconnaisance, DDOS | AndroZoo, BoT-IoT |
| [35] | LSSVM | Anomaly, Malware | Drebin KDD99 |
| [36] | DFEL | Dataset Multiple | UNSW-NB15, |
| [33] | LSTM | DoS Flooding | ISCX2012, |
| [37] | Adaboost | Botnet Flooding | DOS, UNSW-NB15 |

## III. Literature Review

The article by Syed Rizvi et al. [41] presents a method for identifying and recording potential malicious attacks in real time with minimal overhead, making it suitable for constrained environments. They used the IoT-23 dataset for their experiment. The results showed that each of the algorithms used achieved a classification accuracy of over 99% on a representative device with limited resources. The model was deployed on a Raspberry Pi 1, using a desktop computer equipped with an i7-1165G7 processor and 8GB RAM. However, it's worth mentioning that the concept of a resource-constrained device has changed over time. For example, modern Raspberry Pi models now offer multiple gigabytes of RAM and a quad-core CPU. This shift blurs the distinction between traditional resource-constrained devices and more powerful computing platforms. As a result, it opens up new possibilities for what these devices can achieve, expanding their capabilities beyond previous limitations.

In another study, Selim Ylmaz et al. [41] introduced a transfer learning framework for intrusion detection in IoT networks. This framework consists of three components: a feature extractor, a knowledge base, and a classifier. They proposed two transfer learning scenarios for IoT security: transferring knowledge for new devices and transferring knowledge for new attacks1. However, this study did not implement the methodology on real resource-constrained hardware, which could have validated the results and identified potential difficulties arising from the lack of resources.

Arshad et al. [42] conducted research illustrating how the COLIDE framework could be used as a collaborative intrusion detection system in resource-constrained IoT devices. They used CONTIKI OS to simulate an IoT network environment without a malicious node. However, this setup does not include a malicious entity attacking the devices or any on-board detection methods on the simulated devices. Even though the paper claims to have achieved a high detection performance, it provides no metrics to demonstrate its accuracy and f1-score, and does not discuss the computational and energy resources required by their models.

Vittorio Cozzolino et al. [43] tackled the security and privacy concerns stemming from the widespread use of resource-constrained IoT devices. Notably, they observed that manufacturers often prioritise rapid market entry over implementing robust security measures due to fierce competition. Additionally, existing security tools are ill-suited for IoT devices. To address these issues, the researchers developed a lightweight intrusion detection system tailored specifically for resource-constrained IoT devices. They crafted a prototype using the IncludeOS unikernel, ensuring minimal resource usage and a streamlined codebase. Their evaluation, conducted on both x86 and ARM devices, included comparisons with Snort, a popular network intrusion detection system. Importantly, they tested their prototype on single-board computers, rather than microcontrollers which are prevalent in the IoT landscape. Furthermore, their experimental setup involved running the IDS in a virtualised environment, introducing additional processing overhead to the machines. Despite these challenges, their results demonstrated that the prototype effectively detected attack patterns while consuming significantly fewer CPU and RAM resources compared to Snort, offering a promising solution for securing resource-constrained IoT devices.

Ananthi, P et al. [44] suggest utilising Recursive Feature Elimination (REE) with the KDD 99 dataset to construct an Intrusion Detection System (IDS) for IoT devices. REE is an algorithm that selects the most relevant features by iteratively removing irrelevant or redundant ones until the optimal subset is attained. The KDD 99 dataset contains network traffic data used for training a deep neural network to classify normal and malicious network traffic. The performance of the proposed IDS is evaluated using metrics such as recall, precision, F1-score, and accuracy. Additionally, hyperparameter tuning and ensemble learning techniques are employed to improve the IDS's performance. What this research lacks is new fresh data from up-to-date machines. The KDD99 dataset is obsolete by today's standards and is only valuable in benchmarking new models. The dataset was created based on network traffic collected in a controlled environment, which may not reflect the complexities and diversity of contemporary network environments.

Hichem Sedjelmaci et al. [45] highlighted the importance of striking a balance between detection accuracy and energy consumption, particularly concerning low-resource IoT devices where continuous activation of anomaly detection could lead to excessive energy usage. Through the application of game theory and Nash equilibrium, the authors introduce a novel approach where anomaly detection is selectively activated based on the expectation of new attack signatures, thus optimising energy consumption while maintaining high detection accuracy. Simulation results presented in the paper illustrate the effectiveness of this approach, demonstrating both low energy consumption and high detection rates with minimal false positives. Just like previous papers, this paper focuses on running the model in a simulated environment and not on real hardware. Even though the software used to simulate the example was mentioned there are no hardware specifications which would be crucial in identifying any potential limitations or issues as results would vary from platform to platform.

Our research sets itself apart by implementing an intrusion detection system on a real, physically resource-constrained device that is widely used in both homes and industries. We also examine its flaws and power consumption and validate its feasibility as a potential solution to RC-IoT security. We also look at potential future solutions in expanding this research to take into account the existing limitations of these devices.

## IV. Experimental Setup for Network Anomaly Detection

This section outlines our research methodology, covering the experimental design, choice of microcontroller, and the role of ESP32 in our setup. It explains how we conducted our experiments, why we made certain decisions and the challenges of implementing network anomaly detection on IoT devices. Serving as a practical guide to our research process, this section focuses on the hands-on aspects of our work.

### A. Microcontroller of choice

In the realm of the IoT, the choice of microcontroller plays a pivotal role in determining the efficiency and effectiveness of the system. The ESP series, developed by Espressif Systems, offers a range of microcontrollers that are widely used in IoT applications due to their integrated Wi-Fi and Bluetooth capabilities, compact design, and cost-effectiveness. The most prominent microcontrollers in commercial IoT devices are Atmel, STM, ESP8266, and ESP32 [46]. These microcontrollers are chosen based on their features, such as the number of bits, architecture, and memory [46].

When considering the price-to-performance ratio, the ESP32 series offers more advanced features and better performance compared to the ESP8266. For instance, the ESP32 includes a dual-core Xtensa 32-bit LX6 microcontroller, while the ESP32-S2 and ESP32-S3 offer a single-core and dual-core Xtensa 32-bit LX7 microcontroller respectively [47]. The ESP32-C3, on the other hand, is equipped with a single-core RISC-V microcontroller [47]. Concerning table III in terms of library maturity and stability, the ESP8266 has a well-established Arduino core, which allows developers to write sketches using familiar Arduino functions and libraries. The ESP32 and its variants (ESP32-S2, ESP32-S3, and ESP32-C3) also have robust Arduino cores and are supported by Espressif's open-source ESP-IDF, which provides a rich set of libraries and system features. The ESP-IDF is continually updated, indicating a high level of library maturity and stability [48].

In conclusion, the ESP32-WROOM, which utilises the ESP32-S3 chip, emerges as the optimal choice for IoT applications that require robust performance, particularly in the realm of Machine Learning and AI. The dual-core architecture of the ESP32-S3, based on the ARM chipset, provides superior computational power, which is a critical factor for processing-intensive tasks inherent in Machine Learning and AI applications. Furthermore, the maturity and stability of ESP32-S3, as evidenced by its extensive use and support in the developer community, further reinforce its suitability for complex IoT systems [49]. This is in contrast to the newer RISC-V-based variants, which, while promising, may not yet offer the same level of stability and community support. Therefore, for applications that demand high performance, reliability, and a mature and stable development platform, the ESP32-WROOM, with its ESP32-S3 chip, presents a compelling choice. However, it is important to note that the choice of microcontroller should always be guided by the specific requirements and constraints of the individual IoT application.

| | **ESP8266** | **ESP32** | **ESP32-S2** | **ESP32-S3** | **ESP32-C3** | **ESP32-C6** |
|---|---|---|---|---|---|---|
| Announcement Date | 2014, August | 2016, September | 2019, September | 2020, December | 2020, November | 2021, April |
| Main processor | Tensilica L106 32-bit | Tensilica Xtensa 32-bit LX6 | Tensilica Xtensa 32-bit LX7 | Tensilica Xtensa 32-bit LX7 dual-core | RISC-V 32-bit | RISC-V 32-bit |
| SRAM | 160KB | 520KB | 320KB | 512KB | 400KB | 400KB |
| ROM | 0 | 448KB | 128KB | 384KB | 384KB | 384KB |
| JTAG | ✗ | ✓ | ✗ | ? | ✓ | ✓ |
| Cache | 32 KB instruction | 64KB | 8/16KB (configurable) | ? | 16KB | ? |
| WiFi | Wi-Fi 4 (only up to 72.2Mbps) | Wi-Fi 4 | Wi-Fi 4 | Wi-Fi 4 | Wi-Fi 4 | Wi-Fi 6 |
| Bluetooth | X | BLE 4.2 (upgrade to 5.0, with limitations) | X | BLE 5.0 | BLE 5.0 | BLE 5.0 |
| Ethernet | X | ✓ | X | ? | X | ? |
| RTC memory | 768B | 16KB | 16KB | 16KB | 8KB | ? |
| PMU | ✓ | ✓ | ✓ | ? | ✓ | ? |
| ULP coprocessor | X | ✓ | ULP-RISC-V | ? | X | ? |
| Cryptographic Accelerator | X | SHA, RSA, AES, RNG | SHA, RSA, AES, RNG, HMAC, Digital Signature | SHA, RSA, AES, RNG, HMAC, Digital Signature | SHA, RSA, AES, RNG, HMAC, Digital Signature | SHA, RSA, AES, RNG, HMAC, Digital Signature |
| Secure boot | X | ✓ | ✓ | ✓ | ✓ | ✓ |
| Flash encryption | X | ✓ | XTS-AES-128/256 | ✓ | XTS-AES-128 | XTS-AES-128 |
| SPI | 2 | 4 | 4 | ? | 3 | ? |
| I2C | 1 | 2 | 2 | ? | 1 | ? |
| I2S | 1 | 2 | 1 | ? | 1 | ? |
| UART | 2 (one TX only) | 3 | 2 | ? | 2 | ? |
| SDIO Host | 0 | 1 | 0 | 2 | 0 | 0 |
| SDIO Slave | 0 | 1 | 0 | 0 | 0 | 0 |
| GPIO | 17 | 34 | 43 | 44 | 22 | 22 |
| LED PWM | 5 | 16 | 8 | ? | 6 | ? |
| MCPWM | 0 | 6 | 0 | 2 | 0 | 0 |
| Pulse counter | 0 | 8 | 4 | ? | 0 | X |
| GDMA* | 0 | 0 | 0 | ? | 6 | ? |
| USB | X | X | USB OTG 1.1 | ? | Serial/JTAG | ? |
| TWAI** | 0 | 1 | 1 | ? | 1 | ? |
| ADC | 1x 10-bit SAR | 2x 12-bit SAR, up to 18 channels | 2x 13-bit SAR, up to 20 channels | ? | 2x 12-bit SAR, up to 6 channels | ? |
| DAC | X | 2x 8-bit | 2x 8-bit | ? | X | X |
| RMT | 1x transmission + 1x reception | 8x transmission/reception | 4x transmission/reception | ? | 2x transmission + 2x reception | ? |
| Timer | 2x 23-bit*** | 4x 64-bit | 4x 64-bit | ? | 2x 54-bit + 1x 52-bit | ? |
| Temperature Sensor | ✓ | ✓ | ✓ | ? | ✓ | ? |
| Hall Sensor | X | ✓ | X | ? | X | ? |
| Touch Sensor | 0 | 10 | 14 | ? | | |

**TABLE III:** Comparison of ESP microcontrollers [50]

*1) ARM vs RISC-V:* ARM and RISC-V are two processor architectures that have gained significant attention in recent years [51]. ARM is a proprietary Instruction Set Architecture (ISA) that has become the dominant choice for embedded systems and mobile devices due to its longstanding presence in the market, as well as years of trust and expertise cultivated resulting in widespread reputation [51]. On the other hand, RISC-V is an open-source ISA based on the Reduced Instruction Set Computing (RISC) principles1. It is licence-free and royalty-free, allowing users to extend the ISA with new instructions and innovate the microarchitecture of RISC-V processors for free [52]. When comparing these two architectures, there are several key differences to consider.

1) Openness: RISC-V is open-source and license-free, which means anyone can use it to design their chips without paying license fees. ARM, on the other hand, is proprietary and controlled by specific companies, offering established reliability and compatibility but limiting customisation [52] [53].
2) Usage: ARM has been the dominant choice for embedded systems and mobile devices due to its long-standing presence on the market. RISC-V, being relatively new, is still gaining traction in the industry [53].
3) Efficiency: Both architectures are efficient in their own way. ARM has been making power-efficient processors for decades [52]. RISC-V, on the other hand, is seen to be flexible and efficient, and its open-source nature allows continuous improvements.

Concluding because of the stability and maturity of the ARM platform we opted to use the ESP32S WROOM module.

### B. Testbed

Our experimental setup comprises several key components. In our testbed, we replicated a typical sensing IoT device using a temperature/humidity sensor connected to a home network, which is a prevalent use case for IoT devices, which is a prevalent use case for IoT devices [54] [55] [56]. This network comprises several key components, each serving a different purpose. An access point is included to ensure network connectivity. A smart device, specifically an ESP32, is incorporated to represent the IoT devices typically found in such networks. To simulate the communication between the device and a hub, an MQTT server is used for message aggregation. MQTT is one of the most popular publication/-subscription services used for IoT [55] ensuring that our test bed is as realistic as possible. By relying on it we ensure that legitimate data is flowing back and forth between the device and the server simulating a real use case. Finally, an attacker is also part of the setup to represent potential security threats in a real-world scenario.

This configuration was chosen for its relevance and realism, as it closely mirrors the conditions under which these devices operate in everyday settings. It allows us to assess the performance and security of the ESP32 in an environment that is both controlled and reflective of real-world conditions. The setup includes a total of four devices, each playing a crucial role in the network. By adopting this setup, our objective is to provide a comprehensive and practical evaluation of the capabilities and vulnerabilities of ESP32, thus contributing valuable insights into the field of IoT security.

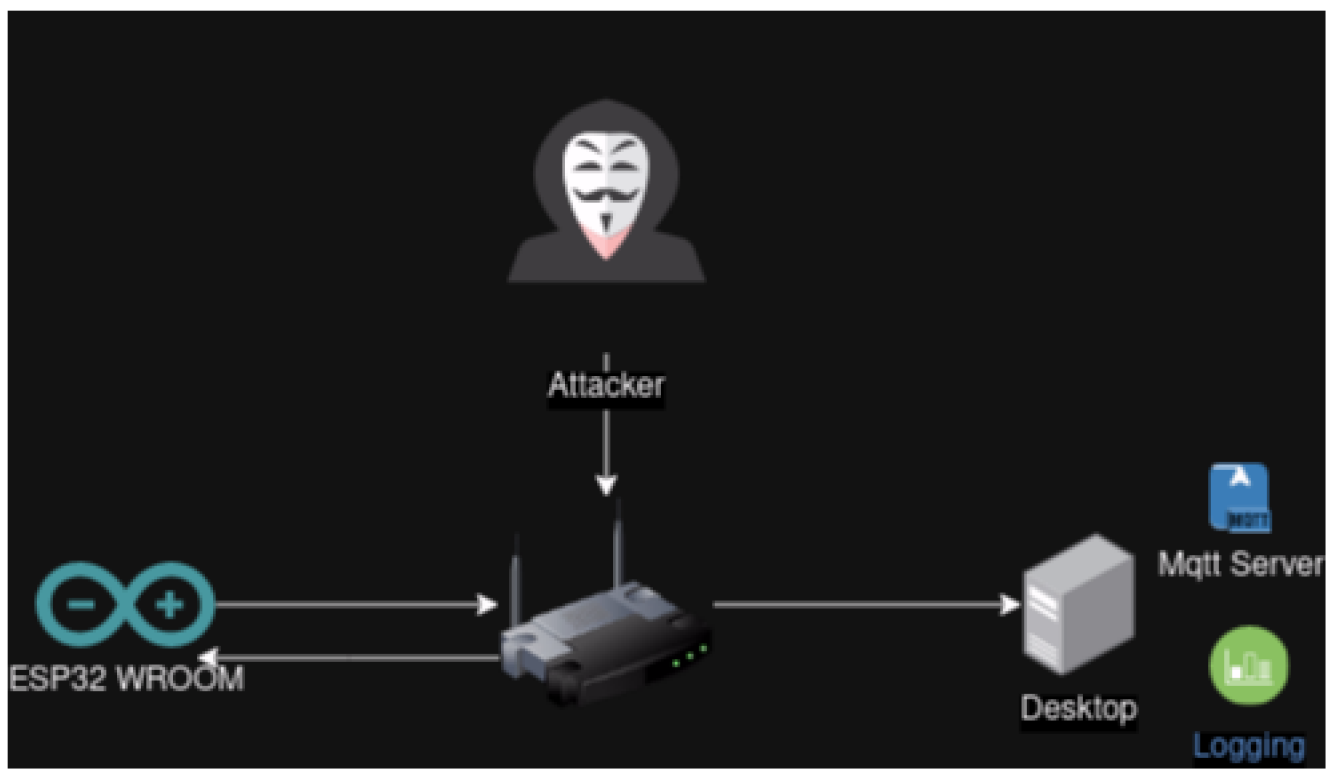

**Fig. 3:** Experimental Setup

*1) ESP32's Role:* The ESP32 microcontroller is a critical component in this setup:

- It sends telemetry data to an MQTT server hosted on the desktop computer. This data includes sensor readings and device status updates.
- The ESP32 captures network traffic data to analyse communication patterns and data exchanges.

*2) Desktop Computer and MQTT Server:* The desktop computer is the control centre of the experiment:

- It hosts an MQTT server to receive telemetry data from the ESP32.
- A serial connection between the desktop and the ESP32 is established to collect captured network traffic data.

*3) Denial of Service (DoS) Attacks on ESP32:* The experimental setup includes subjecting the ESP32 to various DoS attacks, such as:

- UDP and TCP floods: Overwhelming the ESP32's network interface with excessive packets.
- Ping of Death: Sending oversized ICMP packets to destabilize the ESP32.
- SYN floods: Flooding the ESP32 with high-SYN requests to consume its resources.
- Man-in-the-Middle (MITM) Attacks: Intercepting and relaying communication between devices.
- Sniffing: Intercepting and analysing network traffic for sensitive information.

*4) Isolated Network with Access Point:* The TP-LINK AC1200 access point used in our experimental setup connects the ESP32 to an isolated network with multiple devices such as other ESP32 devices, Raspberry PI 3s, acting as simplistic IoT devices taking temperature, humidity and pressure measurements. This introduces network noise and complexity to the experiment, simulating a real-world network scenario.

### C. Data Gathering

In our research, we utilised the Arduino Netdump library, a powerful tool that enables us to monitor the surface-level TCP traffic flow through the ESP32.

While the Netdump library does not provide deep packet inspection, a method that would allow us to examine the details of individual packets, it does offer us the ability to identify patterns in the network traffic. This can be incredibly useful for understanding the general behaviour of the network and identifying potential issues or anomalies.

It is important to note that this method only allows us to view the traffic that is flowing to and from the ESP32 device itself, not the entire network. This means that we are only getting a snapshot of the network activity, rather than a complete picture.

Additionally, we considered the possibility of setting the ESP32 in promiscuous mode. This would allow us to view the "wild" packets in the area, essentially giving us the ability to monitor all network traffic within the device's range. However, enabling promiscuous mode would prevent the device from maintaining a stable connection to the network, which was a requirement for our research. Therefore, we decided that this approach was outside the scope of our current project.

### D. Attacking the device

In our study, we configured the device to collect data over 24 hours. This duration was chosen as it aligns with the time frame in which most commonly used datasets are created, as referenced earlier in Table II. This time frame is significant as it encompasses a full day's cycle of network activity, capturing the variability and patterns that occur throughout the day.

During these 24 hours, cyber-attacks were executed sporadically. This was accomplished by setting up a script to run at random times throughout the day a sample of which is depicted in Figure 4. This approach was taken to mimic the real-world scenario where attacks do not occur continuously but rather at random intervals. By interspersing these attacks throughout the day, we were able to create a realistic dataset where the majority of the traffic is normal, with only a small proportion representing malicious activity.

```
1  #!/bin/bash
2  # Directory containing the scripts
3  SCRIPT_DIR="./scripts"
4  # Get a list of all scripts
5  SCRIPTS=($(ls $SCRIPT_DIR))
6  # Calculate the number of seconds in 24
       hours
7  SECONDS_IN_DAY=86400
8  while true; do
9    # Pick a random script
10   SCRIPT=${SCRIPTS[$RANDOM %
         ${#SCRIPTS[@]}]}
11   bash "$SCRIPT_DIR/$SCRIPT"
12   # Wait for a random amount of time
13   sleep $((RANDOM % SECONDS_IN_DAY))
14 done
```

**Fig. 4:** Bash script for running scripts randomly over a 24-hour period

Considering the nature of the sensor and its consistent behaviour regardless of time (day, night, weekdays, weekends), capturing 24 hours of data is sufficient to represent all its activities. This is also consistent with practices in similar studies, which typically utilise comparable durations of data collection.

Given that the data flowing between the board and the MQTT server primarily consists of sending Wi-Fi signal strength telemetry, there is minimal variation in traffic. Thus, collecting data over longer periods would not provide additional valuable insights and would be redundant.

The sporadic nature of the attacks within the dataset further allows for a thorough examination of how well these systems can detect and respond to threats amidst normal network traffic. The surface-level traffic captured from a single device resulted in a dataset size of 5.1MB.

### E. Results and Evaluation

Data for this study were collected directly from the ESP32 microcontroller, enabling us to customise the model based on its unique feature set. The collected data were then channelled through the serial output of the microcontroller and subsequently exported to a CSV file for further analysis. The structure of the data is as depicted in Table IV, with the target column manually labelled as either 'Legitimate' or 'Anomalous'. Since we already knew when our attacks were deployed, it was a straightforward process to label the data. This labelling process ensured that the model, when trained, would align perfectly with the expected format of the ESP32. This approach guarantees a more accurate and reliable analysis.

To further generalise our model and enhance its applicability across different scenarios, we explicitly removed IP addresses, ports, and timestamps from the dataset. This was done for the following reasons:

- **Removal of IP Addresses and Ports**:
  - *Generalisation*: Including IP addresses and port numbers can make the model overly specific to the network environment in which the data were collected. By excluding these elements, the model becomes more generalised and is better suited to detect anomalies across various network configurations and devices, rather than just the specific setup used during data collection.
  - *Privacy*: Removing IP addresses and ports helps protect the privacy and security of the network environment, ensuring that sensitive information is not inadvertently exposed through the dataset.
- **Exclusion of Timestamps**:
  - *Arduino Time Library Limitation*: The Arduino time library, which the ESP32 relies on, has a known limitation where the counter resets every 50 days. For a device expected to operate 24/7, this reset could introduce inconsistencies and inaccuracies in the data, particularly over long-term monitoring.
  - *Pattern Identification*: While timestamps can be useful for identifying temporal patterns and trends, the

| DIRECTION | IPV | IP | TYPE | PORT | SEQ | ACK | WIN | LEN | target |
|---|---|---|---|---|---|---|---|---|---|
| out | IPv4 | 10.42.0.71 > 10.42.0.153 | TCP | 51795 > 1883[.] | seq:10339 | ack:602218196 | win:35 | | LEGIT |
| in | IPv4 | 10.42.0.153 > 10.42.0.71 | TCP | 1883 > 51795[P.] | seq:602218196..602218226 | ack:10339 | win:64051 | len=30 | LEGIT |

**TABLE IV:** Dataset Structure

potential inaccuracies due to the reset issue make them less reliable in this context. By excluding timestamps, we avoid these pitfalls and ensure that our model focuses on the inherent features of the data that are consistent and reliable.

– *Robustness*: Without relying on timestamps, the model can become more robust to variations in timing, making it applicable in real-time scenarios where exact timing may not be synchronised or available.

In general, these adjustments help create a more versatile and effective anomaly detection model that can be deployed in various settings without being overly dependent on specific network parameters or temporal data, which can be prone to inconsistencies. This ensures that the model is not only tailored to the ESP32's capabilities but is also broadly applicable and reliable in detecting legitimate versus anomalous behaviours across different environments.

### F. Choosing a Model

Given that we are utilising EloquentTinyML for this project, we are constrained by the types of methods available for model training. The reasoning behind using EloquentTinyML is that it allows us to run machine learning algorithms easily on resource-constrained devices without dedicated ML/AI hardware. At the time of writing, only basic neural networks and decision trees can be employed for simple ESP32s that lack floating-point arithmetic coprocessors. This limitation is due to the computational capacity of these devices, which necessitates the use of simpler, less resource-intensive algorithms. Despite these constraints, these methods can still provide valuable insights and contribute to the development of effective models for microcontroller-based IoT devices [57]. While it is possible to perform machine learning tasks without a floating-point coprocessor, the absence of one can limit the precision, performance, and complexity of the models that can be used, and make memory management more challenging [58] [57]. However, advancements in technology and algorithm design are continually pushing these boundaries, making it increasingly feasible to deploy sophisticated machine-learning models on devices with limited resources [58].

### G. Deploying the Neural Network model

Deploying the model on the ESP32 presents several challenges, primarily due to the limited documentation available on model implementation in specific contexts, such as intrusion detection systems, for such devices. Furthermore, given the limited resources of the ESP32, the deployed model must be optimised for both memory usage and computational efficiency. This adds another layer of complexity to the task. Despite these challenges, we successfully deploy a model using a deep neural network architecture with layers of 9 (input layer), 8 (hidden layer 1), 7 (hidden layer 2), 5 (hidden layer 3), 3 (hidden layer 4), 2 (output layer) nodes. This architecture is chosen because it produces the least amount of false positives in our tests. Adding more layers reduces false positives, but the point of diminishing returns appears when a 6th layer is added.

As depicted in Figure 5, the network starts with 9 input layers using a ReLU activation function and finishes off using a sigmoid activation function for classification. As the output is binary, using a sigmoid activation function makes sense. After compiling the model using Python with TensorFlow, it is later converted to pure C code for deployment on the Arduino (Tflite to C). This allows the model to be saved into hex format with a minimal footprint of 26KB, providing the ESP32 board plenty of space for additional functionality. The model is trained using 200 epochs, giving an F1 score of **96%**.

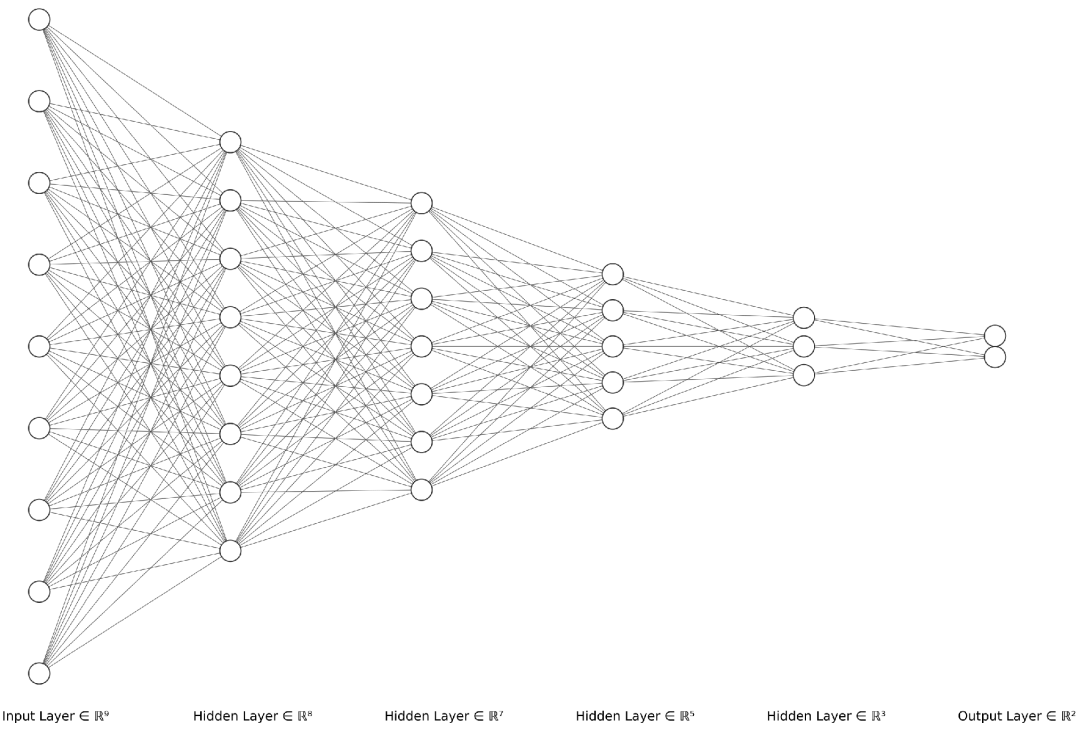

**Fig. 5:** Neural Network Layout

The graph 6 illustrates the model accuracy across 200 epochs for both the training and testing datasets. Several key observations can be made:

Initially, at epoch 0, the accuracy starts at around 0.90 for both datasets, indicating a strong initial performance possibly due to pre-training or initial configuration.

Subsequently, there is a notable increase in accuracy within the initial epochs for both training and testing datasets. The training accuracy quickly approaches near-perfection, stabilising at nearly 1.0, demonstrating efficient learning from the training data.

Around the 10th epoch, both training and testing accuracies stabilize at approximately 0.966, suggesting that the model has reached a consistent state where additional epochs do not significantly improve accuracy. Post-stabilisation, there is minimal fluctuation in accuracy, highlighting consistent performance.

Minor fluctuations in testing accuracy are observed around epochs 10 and 100, albeit very slight. These fluctuations may stem from variations in validation data or adjustments to mitigate overfitting or underfitting.

In comparison, training accuracy consistently remains high and stable, nearing 1.0, while testing accuracy, slightly lower but still robust at around 0.966, maintains a narrow gap. This indicates a well-generalised model with minimal overfitting tendencies.

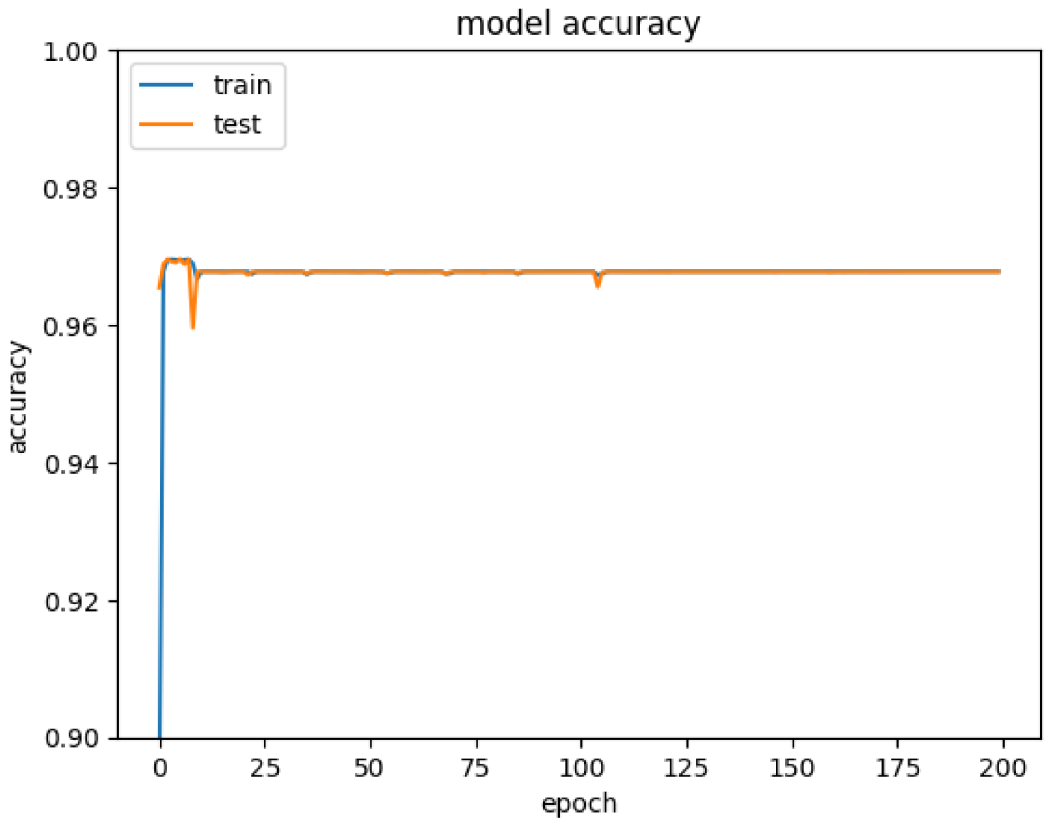

**Fig. 6:** Accuracy over Epochs

### *H. Deploying the Decision Tree*

The development and deployment of the decision tree proved to be more straightforward compared to the neural network. This was primarily because the decision tree was compiled into C code, utilising ready-made functions from standard C libraries, specifically cstdarg. The cstdarg library, also known as "stdarg" in C, is a header in the C and C++ standard libraries that allows functions to accept an indefinite number of arguments. It provides facilities for stepping through a list of function arguments of unknown number and type. This eliminates the need for an external library to be flashed onto the board. However, this comes at a cost of flash storage size as we shall explore in sectionIV-J.

### *I. Performance Enhancement*

Given that the ESP32 features two cores, an advantage over most microcontrollers with only one core, it becomes logical to leverage these dual cores. This enables us to divide the functionalities between them, minimising execution time. Moreover, utilising the proto thread library enables further task distribution through scheduling. This approach allows us to execute multiple functions concurrently. However, it is important to note that using proto-threading without careful consideration could lead to inefficiencies. Utilising a proto thread instead of allocating it to a separate core may result in extended execution times during model operation, consequently impeding the performance of other functions. This issue was evident during a denial of service attack on the device to test our model, where the model's execution dominated the process, preventing data transmission to the MQTT server. To address this, we divided the model to run on Core 0, while the remaining ESP32 functionality operated on Core 1 (telemetry). This is depicted in figure 7 which illustrates how the functionality has been split up.

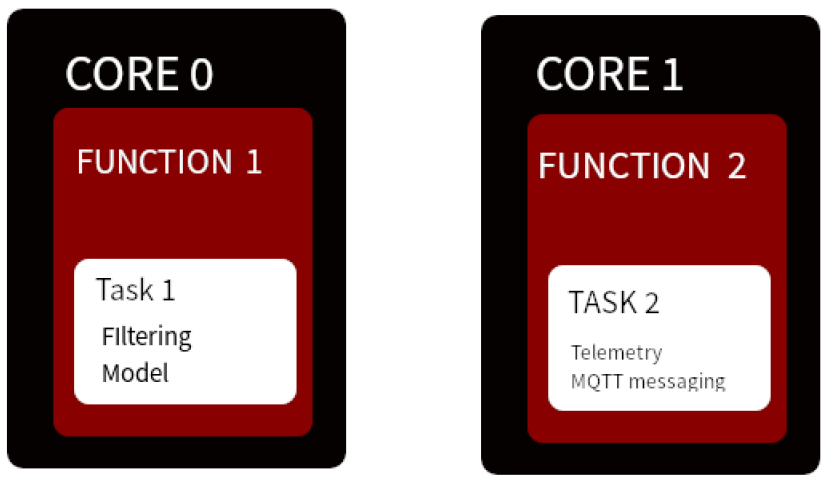

**Fig. 7:** Core Utilisation

Expanding by utilising direct port manipulation, the execution time for turning on the onboard indicator LED when an anomaly was detected was reduced. The LED was used as a visual indicator if the model was detecting an anomaly or not without having to send data to another device thus reducing unwanted network noise. Direct port manipulation is a method that allows for lower-level and faster manipulation of the I/O pins of the microcontroller on an Arduino board. This method can significantly speed up tasks compared to using higher-level functions like "digitalWrite" [59]. High-level functions like "digitalWrite" have additional checks and operations that add overhead. For example, they convert the PIN to the corresponding port and bit, and they turn off interrupts while changing the pin state [59]. Direct port manipulation bypasses these operations, resulting in less overhead and faster execution.

### *J. Model Size comparison*

When comparing the model size of a neural network and a decision tree, it is important to consider not just the size, but also the dependencies and performance of each model. A neural network model might be 10 times smaller than a decision tree model. This smaller size can be advantageous in terms of storage and memory usage, especially on devices with limited resources such as an Arduino. However, this comes with a dependency on an external library to interpret the model and make predictions. This reliance can introduce additional complexity and potential failure points, as the model's performance is tied to the functionality and performance of the library. On the other hand, a decision tree model can run natively on the Arduino without the need for an external library. This can simplify the deployment and operation of the model, as there are fewer dependencies to manage. However, the larger size of the decision tree model could be a disadvantage on devices with limited storage or memory. In summary, the choice between a neural network and a decision tree depends on the specific requirements and constraints of the implementation. If memory and storage are major concerns, a smaller neural network might be more suitable. If simplicity and independence from external libraries are more important, a decision tree could be better.

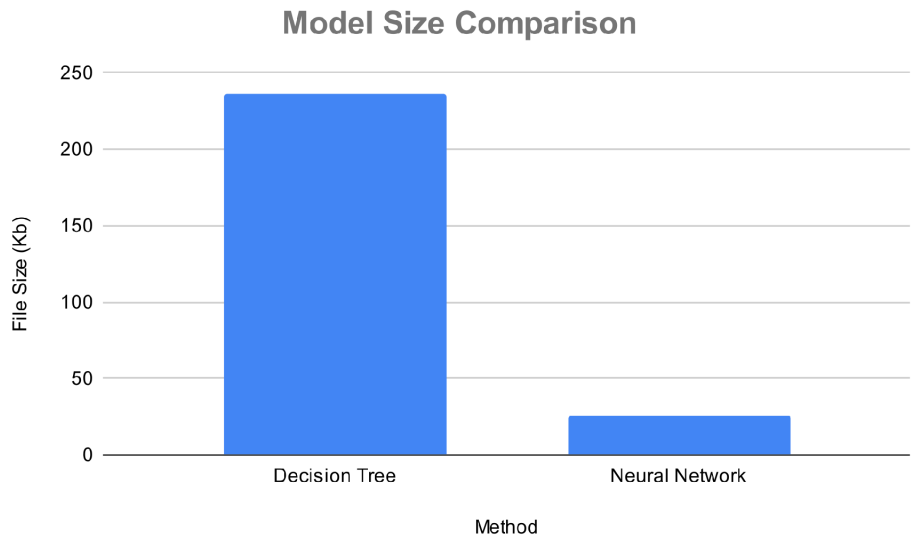

**Fig. 8:** Comparison of model size

### *K. Reducing Power Consumption and Enhancing Efficiency*

Continuous operation of the model can be inefficient, as it may produce multiple false positives. Thus, implementing hard-coded filtering checks on read packets is crucial. These checks can detect abnormal packet rates and corrupted packets, replacing the need to run the model. Using these checks, we initially filter packets using a rule before executing the model. This approach not only reduces false positives but also enhances the detection of easily identifiable packet anomalies while conserving processing cycles and power. When a malformed packet is detected it is read by the board as "type(0x6d)" therefore a simple filter such as the one in Figure 9 is suitable for our use case.

```
1   if (readString.length() >0) {
2     if(readString.indexOf("type(0x6d)")
        >=0) {
3       REG_WRITE(GPIO_OUT_W1TS_REG,
          bitUsed);
4     }
5     readString="";
6   }
```

**Fig. 9:** Simple filter for malformed packet

### *L. Power Consumption Analysis*

A significant number of embedded IoT devices rely on batteries for operation, often deployed in remote locations without direct power sources. Therefore, it is prudent for us to assess the power consumption of our model to estimate the necessary battery capacity for 24-hour operation. While the ESP32 does include power-saving features such as deep sleep mode, we will not consider these features universally, as they depend on the specific case. As these devices can be connected to small solar panels, our focus is on calculating the power required to run the device for 24 hours without an external power source.

Since the power consumption of the ESP32 is thoroughly documented and part of the ESP32 documentation [49], we compared our findings with the manufacturers. We concentrated our efforts on assessing power consumption while using our model. By connecting the ESP32 to an oscilloscope with a precision resistor in line with the voltage input pin, we calculated the voltage drop before and after model execution. Applying Ohm's law, we determined the current. Using these values, the formula $V \times I = P$ allowed us to accurately measure power consumption. To further validate our findings, we also used the onboard readings of the bench power supply that gave the current draw of the device. Expanding we also added a small camera to take readings over time of the analysis. Comparing the two, we could validate if the readings were correct. As we can observe in Figure 10 taking measurements using two devices allowed us to validate our results, so we could be sure that they were accurate. As we can observe from

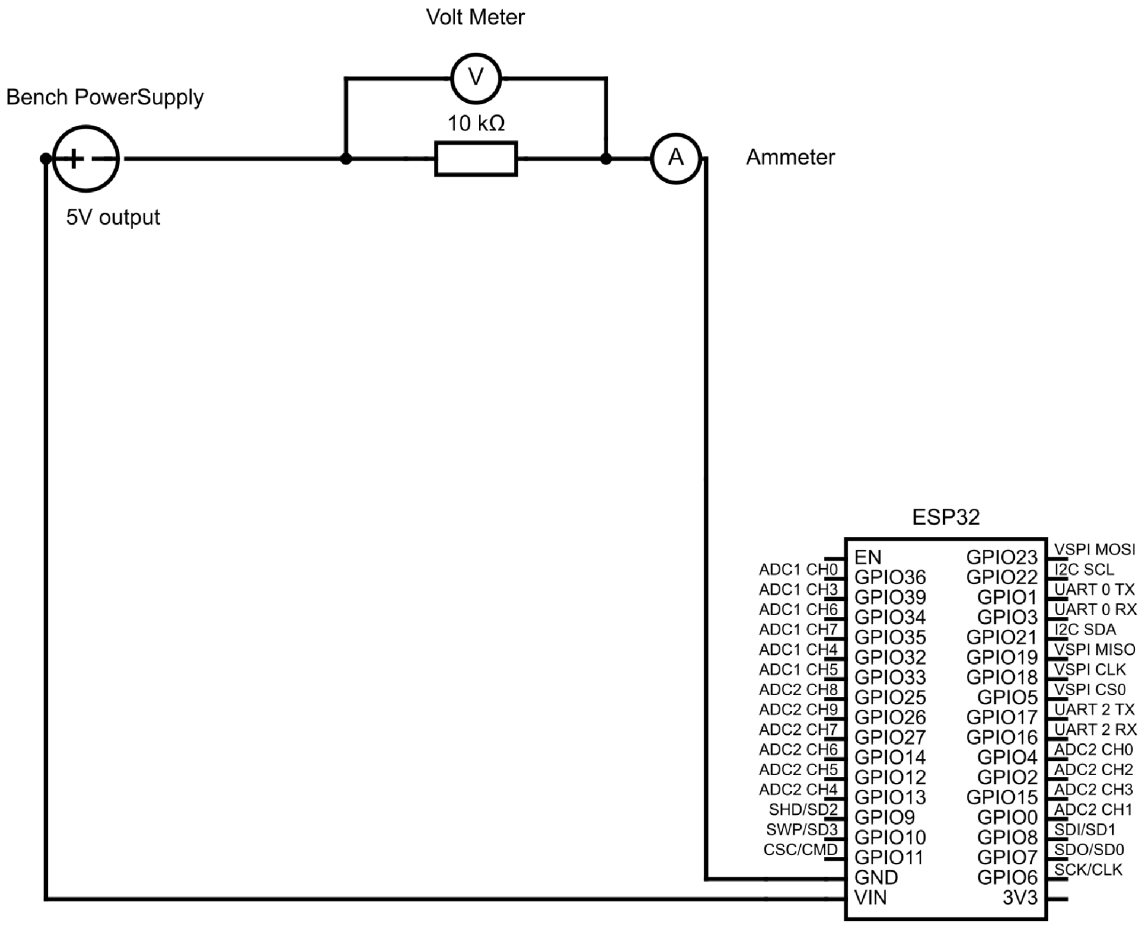

**Fig. 10:** Power Consumption Measurement Setup

Table V and Figure 12 there is predictably more power draw when running the model with the combined filters compared to running without any defence mechanism.

The Figure 11 depicts the power consumption profile of the ESP32 over a 24-hour period. The power consumption varies dynamically, mostly hovering around a baseline of 0.355 watts, representing normal operational levels. However, at random intervals, the power consumption spikes, reaching a maximum of 0.930 watts representing when the model was triggered. Before each peak, there's a gradual increase in power consumption, attributed to a filter power of 0.630 watts. These peaks correspond to periods of increased activity or demand within the device's operation for classifying if traffic is malicious or not. It should be noted that these results will vary

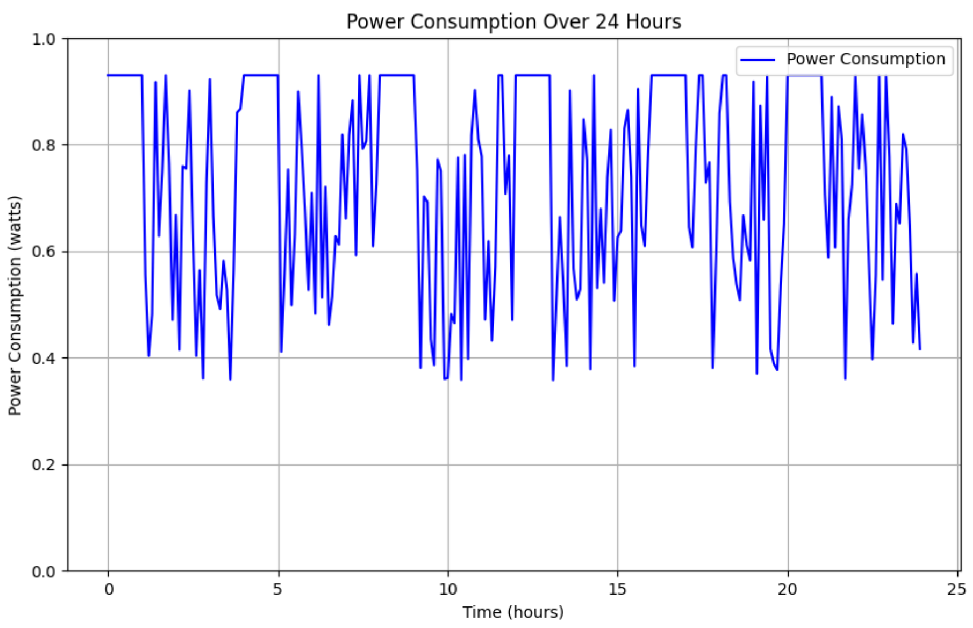

**Fig. 11:** Power Draw over 24 hours

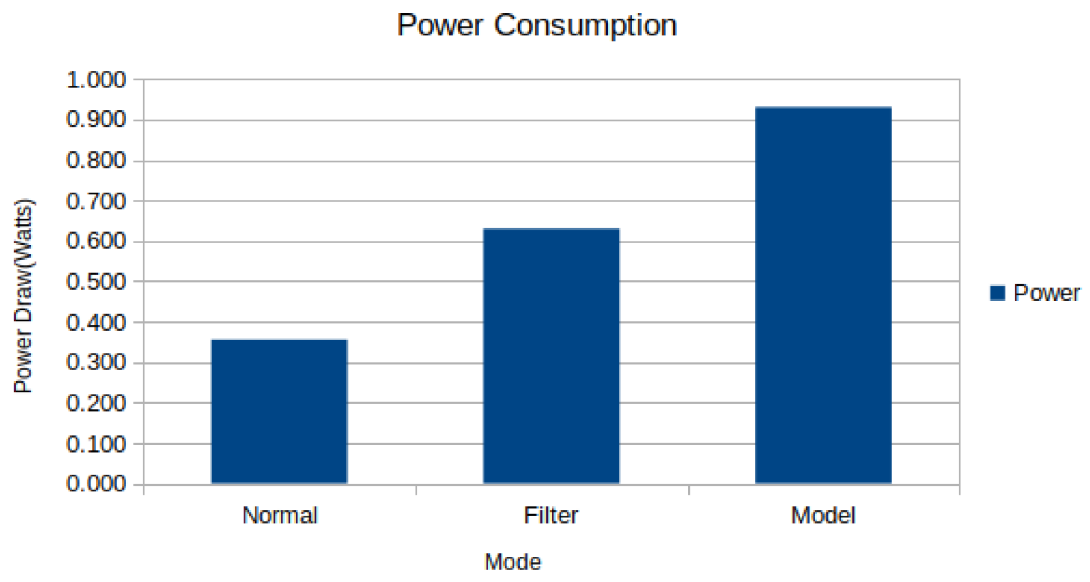

**Fig. 12:** Power Draw of ESP32 in different states

depending on the make and model of the board and cannot guarantee 100% consistency.

| Mode | Power Draw (W) |
|---|---|
| Normal | 0.355 |
| Using Filter | 0.630 |
| Using Model + Filter | 0.930 |

**TABLE V:** Power Consumption

### *M. Battery Capacity Calculation*

To calculate the required battery capacity for lead-acid batteries, the formula is adjusted as follows:

$$B = \frac{100 \times I^k \times t}{100 - q},$$

where:

- $B$ is the battery capacity needed (Ah),
- $I$ is the load current in amperes (A),
- $k$ is Peukert's constant,
- $t$ is the time duration in hours,
- $q$ is the remaining charge as a percentage.

Considering the worst-case scenario of constant 24-hour operation of the model and filters, a battery of **5.6Ah** is required. This is approximately the size of a medium-sized power bank.

*1) Comparison with Manufacturers results:* The manufacturer of the ESP32-S3 microcontroller measures power consumption under various conditions to provide a comprehensive view of the device's energy efficiency. This is typically done using specialised equipment such as a Joulescope ammeter that can accurately measure the current drawn by the device under different modes of operation, such as when the device is idle, under moderate load, or transmitting data [49].

In our tests, we found that the manufacturer's specifications closely match our results validating our experimental findings. The minimum power consumption when the device is idle is similar to the manufacturer's specifications. Furthermore, under moderate load, our measurements were also in line with the manufacturer's data. This consistency reinforces the reliability of the manufacturer's specifications and provides users with a dependable reference for understanding the power efficiency of the ESP32-S3 microcontroller. It should be noted that measurements done by espressif had been undertaken by measuring the chip directly as we measured the development board leading to slightly higher numbers.

| Mode/Condition | Power Consumption |
|---|---|
| Deep sleep | 0.00002W |
| Wake, no WiFi | 0.04W |
| WiFi connected, no traffic | 0.055W |
| Sending/receiving MQTT messages | 0.1W |
| Average power consumption during deep-sleep mode | 0.00002685W |
| Average power consumption during active mode | 0.07832W |
| Total power consumption of one cycle | 0.00637 W |
| Normal current consumption | 0.07W |
| Average current when WiFi connected | 0.12W |

**TABLE VI:** Manufacturers Power Metrics

## V. Limitations of Approach

There are several challenges to implementing an intelligent intrusion detection system on microcontroller-based IoT devices. Firstly, memory constraints on microcontrollers can restrict the size of the dataset used for training and the complexity of the model. This limitation may impact the model's ability to generalise to unseen or evolving attack patterns. Moreover, the storage capacity for updating and maintaining the model's knowledge base may be limited, potentially hindering its adaptability.

Expanding, the dynamic and heterogeneous nature of IoT networks presents challenges in terms of scalability and compatibility. The vast diversity of IoT devices, each with different communication protocols and network architectures, requires careful consideration during the deployment of network intrusion detection solutions. Ensuring seamless integration and compatibility across various IoT devices remains a significant challenge. The vast number of variations of each microcontroller and also the differences in architecture can make it even more difficult to implement a one-size-fits-all approach in a scenario like this. It should also be noted that different communication protocols would pose different overheads leading to vastly difference performance metrics.

In addition to these challenges, overfitting is another significant issue that needs to be addressed. In the context of intrusion detection systems, an overfitted model might perform well on known attack patterns but fail to generalise to new, unseen attacks. This problem is exacerbated by the memory constraints of microcontrollers. A limited dataset might not represent the full range of possible attack patterns, increasing the risk of overfitting. Regularisation techniques, such as L1 and L2 regularisation, can be used to prevent overfitting. However, these techniques add another layer of complexity to the model, which might be challenging given the memory and computational constraints of these devices.

Lastly, the constrained power supply and energy efficiency requirements of microcontrollers pose limitations on continuous monitoring and detection. IoT devices often operate on limited battery power, and running resource-intensive intrusion detection algorithms may drain the battery quickly. Balancing the need for effective intrusion detection with energy efficiency is crucial to ensure the sustainable operation of IoT devices.

Summarising, there are several challenges to implementing an intelligent intrusion detection system on microcontroller-based IoT devices, including memory constraints, scalability and compatibility issues, and energy efficiency requirements. These challenges must be carefully considered when designing and deploying network intrusion detection solutions for IoT devices.

## VI. Conclusion and Future Work

In this research, we unveiled an innovative strategy for deploying network intrusion detection systems on microcontrollers, a vital advancement for the security of IoT devices. Our custom lightweight model, designed for microcontrollers, achieved an impressive F1-score of 96% in detecting diverse network intrusions, improving the security of IoT networks. Furthermore, our decision tree model achieved F1-score of 99%, highlighting the efficacy of our approach. Utilising machine learning techniques and optimising resource consumption, we have proven the practicality of our method on microcontrollers, such as the widely-utilised ESP32. This integration facilitates real-time monitoring and defence of IoT networks, which is crucial for preserving the security and integrity of data communications. However, both models present their respective trade-offs that necessitate careful consideration depending on the use case. Our research also highlights the challenges due to the resource-limited nature of microcontrollers and the complexities of machine learning deployment in IoT. Future endeavours should focus on surmounting these challenges and enhancing the network intrusion detection capabilities for microcontrollers in the IoT domain. While our models have achieved high accuracy, there is always room for improvement. Future work could focus on refining these models to further increase their accuracy and efficiency using different attack scenarios. This could involve exploring different machine learning algorithms, fine-tuning model parameters, or incorporating additional features into the model. IoT networks are dynamic, with devices constantly joining and leaving the network. Future work could explore the development of adaptive models that can learn and adjust to changes in the network environment. As mentioned, both models present their respective trade-offs. A comprehensive analysis could be conducted to understand these trade-offs better and provide guidelines for choosing the most suitable model depending on the use case.

## References

[1] A. Khraisat and A. Alazab, "A critical review of intrusion detection systems in the internet of things: Techniques, deployment strategy, validation strategy, attacks, public datasets and challenges," *Cybersecurity*, vol. 4, no. 1, 2021.

[2] M. Ahmad, "The anatomy of security microcontrollers for iot applications," Jan 2020. [Online]. Available: https://www.digikey.com/en/articles/the-anatomy-of-security-microcontrollers-for-iot-applications

[3] European Parliament and Council, "Regulation (eu) 2016/679 of the european parliament and of the council of 27 april 2016 on the protection of natural persons with regard to the processing of personal data and on the free movement of such data, and repealing directive 95/46/ec (general data protection regulation)," 2016. [Online]. Available: https://eur-lex.europa.eu/legal-content/EN/TXT/?uri=CELEX:32016R0679

[4] A. Rosay, E. Cheval, M. Ghanmi, F. Carlier, and P. Leroux, "Study of network ids in iot devices," *SN Computer Science*, vol. 4, no. 4, 2023.

[5] e. espressif, "Esp32-s3," Dec 2020. [Online]. Available: https://www.espressif.com/en/products/socs/esp32-s3

[6] S. Rizvi, M. Scanlon, J. McGibney, and J. Sheppard, "Deep learning based network intrusion detection system for resource-constrained environments," in *International Conference on Digital Forensics and Cyber Crime*. Springer, 2022, pp. 355–367.

[7] A. Sforzin, F. G. Mármol, M. Conti, and J.-M. Bohli, "Rpids: Raspberry pi ids—a fruitful intrusion detection system for iot," in *2016 Intl IEEE Conferences on Ubiquitous Intelligence & Computing, Advanced and Trusted Computing, Scalable Computing and Communications, Cloud and Big Data Computing, Internet of People, and Smart World Congress (UIC/ATC/ScalCom/CBDCom/IoP/SmartWorld)*. IEEE, 2016, pp. 440–448.

[8] F. B. Insights, "Smart home market size to surpass usd 338.28 billion by 2030, exhibiting a cagr of 20.1https://www.globenewswire.com/en/news-release/2023/07/17/2705500/0/en/Smart-Home-Market-Size-to-Surpass-USD-338-28-billion-by-2030-exhibiting-a-CAGR-of-20-1.html

[9] M. Chui, M. Collins, and M. Patel, "Iot value set to accelerate through 2030: Where and how to capture it," Nov 2021. [Online]. Available: https://www.mckinsey.com/capabilities/mckinsey-digital/our-insights/iot-value-set-to-accelerate-through-2030-where-and-how-to-capture-it

[10] Finite State. (2021) A look back at the top 12 iot exploits of 2021 (part 1). Accessed: 2023-12-05. [Online]. Available: https://finitestate.io/blog/top-12-iot-exploits-of-2021-p1

[11] Threatpost. (2021) Iot attacks skyrocket, doubling in 6 months. Accessed: 2023-12-05. [Online]. Available: https://threatpost.com/iot-attacks-doubling/169224/

[12] IoT World Today. (2021) Iot cyberattacks escalate in 2021, according to kaspersky. Accessed: 2023-12-05. [Online]. Available: https://www.iotworldtoday.com/security/iot-cyberattacks-escalate-in-2021-according-to-kaspersky

[13] Cybersecurity and Infrastructure Security Agency (CISA). (2021) Mitigating log4shell and other log4j-related vulnerabilities. Accessed: 2023-12-05. [Online]. Available: https://www.cisa.gov/news-events/cybersecurity-advisories/aa21-356a

[14] U.S. News World Report. (2023) India's infosys says us unit hit by cyber security event. Accessed: 2023-12-05. [Online]. Available: https://www.usnews.com/news/technology/articles/2023-11-03/indias-infosys-says-us-unit-hit-by-cyber-security-event

[15] Mashable India. (2023) India's biggest data leak: Personal info of 81.5 crore citizens up for sale. Accessed: 2023-12-05. [Online]. Available: https://in.mashable.com/tech/62950/indias-biggest-data-leak-personal-info-of-815-crore-citizens-up-for-sale

[16] The Hindu. (2023) How the personal data of 815 million indians got breached — explained. Accessed: 2023-12-05. [Online]. Available: https://www.thehindu.com/sci-tech/technology/how-the-personal-data-of-815-million-indians-got-breached-explained/article67505760.ece

[17] The Economic Times. (2023) Icmr data leak reveals personal info of 81.5 cr indians, claims report; cbi likely to probe the breach. Accessed: 2023-12-05. [Online]. Available: https://economictimes.indiatimes.com/tech/technology/icmr-data-leak-reveals-personal-info-of-81-5-cr-indians-claims-report-cbi-likely-to-probe-the-breach/videoshow/104865182.cms

[18] U. Utmel, "Using microcontrollers in the internet of things (iot) applications," Jun 2023. [Online]. Available: https://www.utmel.com/blog/categories/microcontrollers/using-microcontrollers-in-the-internet-of-things-iot-applications

[19] S. Kumar, S. Dalal, and V. Dixit, "The osi model: Overview on the seven layers of computer networks," *International Journal of Computer Science and Information Technology Research*, vol. 2, no. 3, pp. 461–466, 2014.

[20] C. Cisco, Oct 2014. [Online]. Available: https://newsroom.cisco.com/c/r/newsroom/en/us/a/y2014/m10/the-internet-of-things-world-forum-unites-industry-leaders-in-chicago-to-accelerate-the-adoption-of-iot-business-models.html

[21] M. Babiuch, P. Foltýnek, and P. Smutný, "Using the esp32 microcontroller for data processing," in *2019 20th International Carpathian Control Conference (ICCC)*. IEEE, 2019, pp. 1–6.

[22] M. Schwartz, *Internet of Things with ESP8266*. Packt Publishing Ltd, 2016.

[23] b. io, "How cyber adversaries attack each of the osi layers 1-7," Feb 2021. [Online]. Available: https://www.byos.io/blog/types-of-cyber-attacks-osi

[24] P. Kumar, Sep 2023. [Online]. Available: https://www.pynetlabs.com/various-kinds-of-osi-layer-attacks/

[25] P. Rawat, "Common security attacks in the osi layer model," Jan 2023. [Online]. Available: https://www.infosectrain.com/blog/common-security-attacks-in-the-osi-layer-model/

[26] A. Alshaibi, M. Al-Ani, A. Al-Azzawi, A. Konev, and A. Shelupanov, "The comparison of cybersecurity datasets," *Data*, vol. 7, no. 2, p. 22, 2022.

[27] S. Prabavathy, K. Sundarakantham, and S. M. Shalinie, "Design of cognitive fog computing for intrusion detection in internet of things," *Journal of Communications and Networks*, vol. 20, no. 3, pp. 291–298, 2018.

[28] C. Liang, B. Shanmugam, S. Azam, M. Jonkman, F. De Boer, and G. Narayansamy, "Intrusion detection system for internet of things based on a machine learning approach," in *2019 International Conference on Vision Towards Emerging Trends in Communication and Networking (ViTECoN)*. IEEE, 2019, pp. 1–6.

[29] P. Illy, G. Kaddoum, C. M. Moreira, K. Kaur, and S. Garg, "Securing fog-to-things environment using intrusion detection system based on ensemble learning," in *2019 IEEE Wireless Communications and Networking Conference (WCNC)*. IEEE, 2019, pp. 1–7.

[30] Z. Tan, A. Jamdagni, X. He, P. Nanda, and R. P. Liu, "A system for denial-of-service attack detection based on multivariate correlation analysis," *IEEE transactions on parallel and distributed systems*, vol. 25, no. 2, pp. 447–456, 2013.

[31] P. Feng, J. Ma, C. Sun, X. Xu, and Y. Ma, "A novel dynamic android malware detection system with ensemble learning," *IEEE Access*, vol. 6, pp. 30 996–31 011, 2018.

[32] S. Fenanir, F. Semchedine, and A. Baadache, "A machine learning-based lightweight intrusion detection system for the internet of things." *Rev. d'Intelligence Artif.*, vol. 33, no. 3, pp. 203–211, 2019.

[33] A. Diro and N. Chilamkurti, "Leveraging lstm networks for attack detection in fog-to-things communications," *IEEE Communications Magazine*, vol. 56, no. 9, pp. 124–130, 2018.

[34] A. Verma and V. Ranga, "Machine learning based intrusion detection systems for iot applications," *Wireless Personal Communications*, vol. 111, no. 4, pp. 2287–2310, 2020.

[35] M. A. Ambusaidi, X. He, P. Nanda, and Z. Tan, "Building an intrusion detection system using a filter-based feature selection algorithm," *IEEE Transactions on Computers*, vol. 65, no. 10, pp. 2986–2998, 2016.

[36] Y. Zhou, M. Han, L. Liu, J. S. He, and Y. Wang, "Deep learning approach for cyberattack detection," *IEEE INFOCOM 2018 - IEEE Conference on Computer Communications Workshops (INFOCOM WKSHPS)*, Apr 2018.

[37] N. Moustafa, B. Turnbull, and K.-K. R. Choo, "An ensemble intrusion detection technique based on proposed statistical flow features for protecting network traffic of internet of things," *IEEE Internet of Things Journal*, vol. 6, no. 3, pp. 4815–4830, 2019.

[38] D. McMillen, "Internet of threats: Iot botnets drive surge in network attacks," Apr 2021. [Online]. Available: https://securityintelligence.com/posts/internet-of-threats-iot-Botnets-network-attacks/

[39] X. He, E. Bauman, F. Li, L. Yu, L. Li, B. Liu, A. Piao, K. W. Hamlen, W. Huo, and W. Zou, "Exploiting the trust between boundaries: Discovering memory corruptions in printers via driver-assisted testing," in *The 21st ACM SIGPLAN/SIGBED Conference on Languages, Compilers, and Tools for Embedded Systems*, ser. LCTES '20. New York, NY, USA: Association for Computing Machinery, 2020, p. 74–84. [Online]. Available: https://doi.org/10.1145/3372799.3394363

[40] A. Rostami, M. Vigren, S. Raza, and B. Brown, "Being hacked: Understanding victims' experiences of IoT hacking," in *Eighteenth Symposium on Usable Privacy and Security (SOUPS 2022)*. Boston, MA: USENIX Association, Aug. 2022, pp. 613–631. [Online]. Available: https://www.usenix.org/conference/soups2022/presentation/rostami

[41] S. Rizvi, M. Scanlon, J. McGibney, and J. Sheppard, "An evaluation of ai-based network intrusion detection in resource-constrained environments," in *2023 IEEE 14th Annual Ubiquitous Computing, Electronics Mobile Communication Conference (UEMCON)*, 2023, pp. 0275–0282.

[42] S. Yılmaz, E. Aydogan, and S. Sen, "A transfer learning approach for securing resource-constrained iot devices," *IEEE Transactions on Information Forensics and Security*, vol. 16, pp. 4405–4418, 2021.

[43] V. Cozzolino, N. Schwellnus, J. Ott, and A. Y. Ding, "Uids: Unikernel-based intrusion detection system for the internet of things," in *DISS 2020-Workshop on Decentralized IoT Systems and Security*, 2020.

[44] P. Ananthi, T. Ramya, and R. Janani, "Ensemble based intrusion detection system for iot device," in *2023 International Conference on Sustainable Computing and Smart Systems (ICSCSS)*, 2023, pp. 1073–1078.

[45] H. Sedjelmaci, S. M. Senouci, and M. Al-Bahri, "A lightweight anomaly detection technique for low-resource iot devices: A game-theoretic methodology," in *2016 IEEE International Conference on Communications (ICC)*, 2016, pp. 1–6.

[46] P. Viorel and C. Mahler, "A guide for selecting the right microcontroller for your iot project," Jun 2023. [Online]. Available: https://iiot-world.com/industrial-iot/connected-industry/a-guide-for-selecting-the-right-microcontroller-for-your-iot-project/

[47] e. espressif, "Get started," Sep 2023. [Online]. Available: https://docs.espressif.com/projects/esp-idf/en/latest/esp32s2/get-started/index.html

[48] ——, "Libraries and frameworks," Jan 2023. [Online]. Available: https://docs.espressif.com/projects/esp-idf/en/latest/esp32s2/libraries-and-frameworks/index.html

[49] ——, "Get started," Nov 2023. [Online]. Available: https://docs.espressif.com/projects/esp-idf/en/latest/esp32s3/get-started/index.html

[50] F. Riccardi, "Comparison table for esp8266/esp32/esp32-s2/esp32-s3/esp32-c3/esp32-c6," Apr 2021. [Online]. Available: https://gist.github.com/fabianoriccardi/cbb474c94a8659209e61e3194b20eb61

[51] S. Sharma, "Risc-v vs arm: A comprehensive comparison of processor architectures," Aug 2023. [Online]. Available: https://www.wevolver.com/article/risc-v-vs-arm-a-comprehensive-comparison-of-processor-architectures

[52] P. Jamieson, H. Le, N. Martin, T. McGrew, Y. Qian, E. Schonauer, A. Ehret, and M. A. Kinsy, "Computer engineering education experiences with risc-v architecturesmdash;from computer architecture to microcontrollers," *Journal of Low Power Electronics and Applications*, vol. 12, no. 3, 2022. [Online]. Available: https://www.mdpi.com/2079-9268/12/3/45

[53] Y. Liu, K. Ye, and C.-Z. Xu, "Performance evaluation of various risc processor systems: A case study on arm, mips and risc-v," in *Cloud Computing – CLOUD 2021*, K. Ye and L.-J. Zhang, Eds. Cham: Springer International Publishing, 2022, pp. 61–74.

[54] S. Kraijak and P. Tuwanut, "A survey on iot architectures, protocols, applications, security, privacy, real-world implementation and future trends," in *11th international conference on wireless communications, networking and mobile computing (WiCOM 2015)*. IET, 2015, pp. 1–6.

[55] F. YALÇINKAYA, H. AYDİLEK, M. Y. ERTEN, and N. İNANÇ, "Iot based smart home testbed using mqtt communication protocol," *International Journal of Engineering Research and Development*, vol. 12, no. 1, p. 317–324, 2020.

[56] M. M. Yamin, B. Katt, E. Torseth, V. Gkioulos, and S. J. Kowalski, "Make it and break it: An iot smart home testbed case study," in *Proceedings of the 2nd International Symposium on Computer Science and Intelligent Control*, ser. ISCSIC '18. New York, NY, USA: Association for Computing Machinery, 2018. [Online]. Available: https://doi.org/10.1145/3284557.3284743

[57] M. Merenda, C. Porcaro, and D. Iero, "Edge machine learning for ai-enabled iot devices: A review," *Sensors*, vol. 20, no. 9, 2020. [Online]. Available: https://www.mdpi.com/1424-8220/20/9/2533

[58] A. Harish, S. Jhawar, B. S. Anisha, and P. Ramakanth Kumar, "Implementing machine learning on edge devices with limited working memory," in *Inventive Communication and Computational Technologies*, G. Ranganathan, J. Chen, and Á. Rocha, Eds. Singapore: Springer Singapore, 2020, pp. 1255–1261.

[59] T. A. Team, "Arduino - portmanipulation," Nov 2023. [Online]. Available: https://docs.arduino.cc/hacking/software/PortManipulation